\documentclass[12pt,thmsa]{article}

\usepackage{cite}

\begin{document}

\title{Accurate eigenvalues of bounded oscillators}
\author{Francisco M. Fern\'{a}ndez \thanks{e-mail: fernande@quimica.unlp.edu.ar}\\
INIFTA (UNLP,CCT La Plata-CONICET), Divisi\'on Qu\'imica Te\'orica,\\
Diag 113 S/N,  Sucursal 4, Casilla de Correo 16,\\
1900 La Plata, Argentina}
\maketitle

\begin{abstract}
We calculate accurate eigenvalues of a bounded oscillator by means of the
Riccati--Pad\'e method that is based on a rational approximation to a
regularized logarithmic derivative of the wavefunction. Sequences of
roots of Hankel determinants approach the model eigenvalues from below
with remarkable convergence rate.
\end{abstract}

\section{Introduction \label{sec:intro}}

Some time ago Fern\'{a}ndez et al \cite
{FMT89a,FMT89b,F92,FG93,F95,F95b,F95c,F96,F96b,F97,F08} developed a method for
the accurate calculation of eigenfunctions and eigenvalues for bound states
and resonances of separable quantum--mechanical models. This
approach is based on Pad\'{e} approximants built from the Taylor expansion
of a regularized logarithmic derivative of the eigenfunction. If such
rational approximations are forced to yield an additional coefficient of the
Taylor expansion, then the physical eigenvalue is given by a convergent sequence of
roots of Hankel determinants constructed from the coefficients of that
series. One merit of this approach, called Riccati--Pad\'{e} method (RPM),
is the great convergence rate of such sequences in most cases, and that the same equation
applies to bound states and resonances\cite
{FMT89a,FMT89b,F92,FG93,F95,F95b,F95c,F96,F96b,F97,F08}. Besides, in some cases
the sequences yield upper and lower bounds to the eigenvalues\cite{FMT89b}.

The approach just mentioned was mainly applied to quantum--mechanical models
with asymptotic boundary conditions at infinity\cite
{FMT89a,FMT89b,F92,FG93,F95,F95b,F95c,F96,F96b,F97,F08}, but it was suggested
that it may be suitable for bounded problems as well\cite{F96}. In this
paper we try the RPM on the bounded anharmonic oscillator recently discussed
by Ciftci and Ateser\cite{CA07} that may be suitable for the study of
quarkonium physics.

In Sec.~\ref{sec:method} we outline the main features of the RPM. In
Sec.~\ref{sec:model} we apply the RPM to the bounded oscillator proposed by
Ciftci and Ateser\cite{CA07}. In Sec.~\ref{sec:spurious} we briefly address
the problem of spurious roots. In Sec.~\ref{sec:central} we show that the RPM
applies to the central--field version of the bounded oscillator. Finally, in
Sec.~\ref{sec:conclusions} we draw some conclusions.

\section{Method \label{sec:method}}

In this section we outline the main features of the Riccati--Pad\'{e} method
(RPM). For concreteness we consider the dimensionless Schr\"{o}dinger
equation with an even--parity potential $V(x)$%
\begin{equation}
\psi ^{\prime \prime }(x)+[E-V(x)]\psi (x)=0  \label{eq:Schrodinger}
\end{equation}
with yet unspecified boundary conditions. We assume that the Taylor
expansion
\begin{equation}
V(x)=\sum_{j=0}^{\infty }V_{j}x^{2j}  \label{eq:V_series}
\end{equation}
converges in a neighbourhood of $x=0$.

The modified logarithmic derivative $f(x)=s/x-\psi ^{\prime }(x)/\psi (x)$
satisfies the Riccati equation
\begin{equation}
f^{\prime }(x)+\frac{2s}{x}f(x)-f(x)^{2}-\frac{s(s-1)}{x^{2}}+V(x)-E=0
\label{eq:Riccati}
\end{equation}
where the fourth term disappears when $s=0$ or $s=1$ that correspond to even
or odd states $\psi (x)$, respectively. The Taylor expansion
\begin{equation}
f(x)=x\sum_{j=0}^{\infty }f_{j}x^{2j}  \label{eq:f_series}
\end{equation}
also converges in a neighbourhood of $x=0$ and the coefficients $f_{j}$
depend on the eigenvalue $E$.

The Pad\'{e} approximant
\begin{equation}
\lbrack N+d/N]=x\frac{\sum_{j=0}^{N+d}a_{j}x^{2j}}{\sum_{j=0}^{N}b_{j}x^{2j}}
\label{eq:Pade}
\end{equation}
yields exactly the first $2N+d+1$ coefficients of the series (\ref
{eq:f_series}). We look for the values of $E$ that allow the Pad\'{e}
approximant to produce the next coefficient $f_{2N+d+1}$ exactly. Such
energy values are given by the roots of the Hankel determinant $H_{D}^{d}=0$
with matrix elements $f_{d+i+j-1}$, $i,j=1,2,\ldots ,D$, where $D=N+1$ is
the matrix dimension, and $d=0,1,\ldots $ is the displacement\cite
{FMT89a,FMT89b,F92,FG93,F95,F95b,F95c,F96,F96b,F97,F08}.

Each Hankel determinant exhibits many roots but one easily identifies
convergent sequences $E_{n}^{[D]}$, $D=2,3,\ldots $ that approach the actual
eigenvalues $E_{n}$ of the Schr\"{o}dinger equation~(\ref{eq:Schrodinger}).
If one such sequence is monotonously increasing\ (decreasing), then each
element provides a lower (upper) bound to the eigenvalue. This property
of the Hankel sequences was already proved for some problems\cite{FMT89b}.

Another interesting feature of the RPM is that one does not have to take
into account the boundary conditions explicitly. The method selects the
appropriate values of $E$ according to the singularities of the potential as
we shall see below.

\section{The model \label{sec:model}}

Ciftci and Ateser\cite{CA07} studied the Schr\"odinger equation~(\ref{eq:Schrodinger})
with the potential
\begin{equation}
V(x)=\frac{a^{2}x^{2}}{\left( 1-x^{2}/R^{2}\right) ^{2}}  \label{eq:V_BO}
\end{equation}
where $a,R>0$, $-R<x<R$, and the boundary conditions $\psi (\pm R)=0$ which are obviously
determined by the poles of $V(x)$. We begin our analysis by noticing that the change of
variables $x=Lq$ leads to
\begin{equation}
\Phi ^{\prime \prime }(q)+\left[ L^{2}E-\frac{a^{2}L^{4}q^{2}}{\left(
1-L^{2}q^{2}/R^{2}\right) ^{2}}\right] \Phi (q)=0  \label{eq:SchrBOtrans}
\end{equation}
where $\Phi(q)=\psi(Lq)$. Therefore, it is clear that one can choose
either $a$ or $R$ equal to unity
without loss of generality as follows from the obvious equalities
$E(a,R)=R^{-2}E(aR^{2},1)=a E(1,aR)$. From a purely numerical point of
view it is convenient to calculate $R^{2}E(1,R)$ for small $R$ and
$E(1,R)$ for large $R$ because $\lim_{R\rightarrow
0}R^{2}E_{n}(1,R)=(n+1)^{2}\pi ^{2}/4$ and $\lim_{R\rightarrow \infty
}E_{n}(1,R)=2n+1$, where $n=0,1,\ldots $ (from now on it should be assumed
that $a=1$ unless otherwise stated).

We first show that the RPM yields the exact harmonic oscillator eigenvalues
when $R\rightarrow \infty $. For example, the determinant of smallest
dimension for $s=0$ and $d=0$ is
\begin{equation}
H_{2}^{0}=\frac{%
E^{6}R^{4}-27E^{4}R^{4}+324E^{3}R^{2}+51E^{2}R^{4}-675E^{2}-324ER^{2}-25R^{4}-81%
}{4725R^{4}}
\end{equation}
so that
\begin{equation}
\lim_{R\rightarrow \infty }H_{2}^{0}=\frac{(E^{2}-25)(E^{2}-1)^{2}}{4725}
\label{eq:H_2^0(HO)}
\end{equation}
which clearly shows that two roots of the Hankel determinant give exactly the
eigenvalues for the first two states of even parity of the harmonic oscillator.

On the other hand, present implementation of the RPM is unable to yield exact
eigenvalues for
$R<\infty $, and the convergence rate of the sequence of roots of the Hankel
determinants decreases as $R$ decreases. However, the RPM provides remarkably
accurate results for all $R$ as shown in what follows.

Table~\ref{tab:table1} shows two
sequences of roots of the Hankel determinants that converge towards the
ground--state eigenvalue of the bounded oscillator given as
$R^{2}E_{0}(1,R),$ for $R=0.1$. The first ten digits of present result
agree with those reported by Ciftci and Ateser\cite{CA07}. The velocity of
convergence of the RPM sequences is as remarkable as in previous
applications of the method to problems with asymptotic boundary conditions
at infinity\cite{FMT89a,FMT89b,F92,FG93,F95,F95b,F95c,F96,F96b,F97,F08}.

We have decided to keep 20 digits in all our results although in some cases
not all of them are exact. In this way one can clearly estimate the accuracy
of an entry by its agreement with the neighbours in the same column and row,
making it easier to appreciate the appearance of new stable digits as $D$ or
$d$ increases.

Table~\ref{tab:table2} shows the ground--state energy for some values of $R$
and the smallest determinant dimension at which all the reported digits are stable
($d=0$ in all cases).

The convergence rate of the RPM sequence decreases as the quantum number
increases. The reason is that $N$ should increase in order to take into
account the increasing oscillation of the excited states. Table~\ref
{tab:table3} shows two RPM sequences converging towards the eigenvalue $%
E_{2} $ for $R=1$. The first sixt digits of our result agree with those
reported by Ciftci and Ateser\cite{CA07}.

\section{Spurious roots \label{sec:spurious}}

As said above, the Hankel determinants exhibit many roots and one has to
select those corresponding to the chosen state and model. Equation (\ref
{eq:H_2^0(HO)}) shows that there is a double root corresponding to the
ground state $E_{0}=1$, and a single root corresponding to the next higher
even state $E_{2}=5$. Multiple roots in an exactly solvable problem become
multiple sequences converging to the eigenvalue of a nontrivial problem\cite
{FMT89a,FMT89b,F92,FG93,F95,F95b,F95c,F96,F96b,F97,F08}.
However, one easily identifies the sequence with the best convergence rate
looking for the root of $H_{D+1}^{d}$ closest to the previously chosen root
of $H_{D}^{d}$. As $D$ increases, more roots appear in the neighbourhood of
the actual eigenvalue which is an additional indication of sound convergence.

The change of variables $x\rightarrow ix$ transforms the Hamiltonian operator
with potential $V(x)=\frac{x^2}{(1-x^2/R^2)^2}$ into minus the Hamiltonian
operator with
potential $V_2(x)=\frac{x^2}{(1+x^2/R^2)^2}$. This transformation accounts for the
spurious roots $E=-1$ and $E=-5$ in equation (\ref{eq:H_2^0(HO)})
(for $R \rightarrow \infty$) as discussed earlier in a more general way\cite{F96}.
When $R<\infty$ the potential $V_2(x)$ does not support bound states (the RPM does
no longer select Dirichlet boundary conditions at $\pm R$, and $-\infty<x<\infty$).
In this case one expects resonances with complex values of $E$,
and outgoing or ingoing waves as $x\rightarrow\pm\infty$ \cite{F95}. For example,
when $R=1$ there is
a sequence of Hankel roots that converge towards $E=-0.015565600439810503080$ that
appears to be minus the real part of a resonance of $V_2(x)$. The complex part of
this resonance seems to be quite small. We do not discuss this feature of the RPM
any further because we are mainly interested in bounded oscillators.

\section{Central--field models \label{sec:central}}

Ciftci and Ateser\cite{CA07} mention a possible application of a three--dimensional
bounded oscillator to quarkonium physics. The application of the
RPM to central--field models is straightforward\cite{FMT89a,F92,F95c,F95}.
The Riccati equation for
the radial part of the Schr\"{o}dinger equation becomes
\begin{equation}
f^{\prime }(r)+\frac{2s}{r}f(r)-f(r)^{2}-\frac{s(s-1)}{r^{2}}+V(r)+\frac{%
l(l+1)}{r^{2}}-E=0  \label{eq:Ricati_CF}
\end{equation}
where $f(r)=s/r-\psi ^{\prime }(r)/\psi (r)$ and $\psi (0)=0$. In this case
$l=0,1,\ldots $ is the angular--momentum quantum number and we choose $s=l+1$
in order to remove the centrifugal term and make $f(r)$
regular at origin\cite{FMT89a,F92,F95c,F95}. When $l=0$ the results are
identical to those of the odd
states of the one--dimensional problem that satisfy the same boundary
condition at origin. Table~\ref{tab:table4} shows two Hankel sequences
converging towards $E_{1}$ (or the energy of the ground state of the
central--field model) when $R=1$. The first six digits of our result agree with
those reported by Ciftci and Ateser\cite{CA07}.

It is worth mentioning that the sequences shown in tables
\ref{tab:table1}, \ref{tab:table3},
and \ref{tab:table4} are monotonously increasing, and, therefore, they
provide lower bounds to the corresponding eigenvalues of this particular
model.

\section{Conclusions \label{sec:conclusions}}

We have shown that the RPM gives accurate eigenvalues for some one--dimensional
and central--field bounded oscillators. The approach automatically selects the physical
energies according to the singularities of the potential--energy function.
In addition to enlarging the field
of application of the RPM we have verified that the perturbation variant of
the AIM proposed by Ciftci and Ateser\cite{CA07} yields reasonably accurate
results for most physical purposes. However, the RPM appears to be much
more accurate and straightforward. It does not require factorization of the
asymptotic behaviour of
the wavefunction, and its Hankel sequences converge monotonously from below, providing
lower bounds to the eigenvalues of this particular model. In this respect the RPM is
a suitable complement to the variational methods that render upper bounds.

Besides, the AIM has not yet proved suitable for the calculation of complex
eigenvalues that the RPM yields easily and accurately\cite{F95c,F95,F96,F96b,F08}.
It is worth mentioning that the RPM is not restricted to the Schr\"odinger
equation. We have
recently applied a variant of the RPM, which we may call Pad\'e--Hankel
method, to nonlinear two--point
boundary value problems, obtaining very accurate results for the
unknown parameters in several models of physical interest\cite{AF07}.

\begin{table}[H]
\caption{Ground--state energy of the bounded oscillator times $R^2$ for $R=0.1$}
\label{tab:table1}
\begin{center}
\begin{tabular}{ccc}
\hline
$D$ & $d=0$ & $d=1$ \\ \hline
\multicolumn{1}{r}{3} & \multicolumn{1}{l}{} & \multicolumn{1}{l}{
2.3697606944397752864} \\
\multicolumn{1}{r}{4} & \multicolumn{1}{l}{2.4662533354307466207} &
\multicolumn{1}{l}{2.4674298730367798888} \\
\multicolumn{1}{r}{5} & \multicolumn{1}{l}{2.4674514194162024875} &
\multicolumn{1}{l}{2.4674515378204140996} \\
\multicolumn{1}{r}{6} & \multicolumn{1}{l}{2.4674515390300695979} &
\multicolumn{1}{l}{2.4674515390347723203} \\
\multicolumn{1}{r}{7} & \multicolumn{1}{l}{2.4674515390348029376} &
\multicolumn{1}{l}{2.4674515390348030263} \\
\multicolumn{1}{r}{8} & \multicolumn{1}{l}{2.4674515390348030267} &
\multicolumn{1}{l}{2.4674515390348030267} \\
\multicolumn{1}{r}{9} & \multicolumn{1}{l}{2.4674515390348030267} &
\multicolumn{1}{l}{2.4674515390348030267} \\
\multicolumn{1}{r}{10} & \multicolumn{1}{l}{2.4674515390348030267} &
\multicolumn{1}{l}{2.4674515390348030267} \\ \hline
\end{tabular}
\end{center}
\end{table}

\begin{table}[H]
\caption{Ground--state energy of the bounded oscillator for some values of $%
R $}
\label{tab:table2}
\begin{center}
\begin{tabular}{ccc}
\hline
$R$ & $E_{0}$ & $D$ \\ \hline
\multicolumn{1}{r}{1} & \multicolumn{1}{l}{2.8848849919939971927} &
\multicolumn{1}{l}{7} \\
\multicolumn{1}{r}{10} & \multicolumn{1}{l}{1.0150378624976100592} &
\multicolumn{1}{l}{3} \\
\multicolumn{1}{r}{100} & \multicolumn{1}{l}{1.0001500037503748711} &
\multicolumn{1}{l}{2} \\ \hline
\end{tabular}
\end{center}
\end{table}

\begin{table}[H]
\caption{Energy of the first even excited state for $R=1$}
\label{tab:table3}
\begin{center}
\begin{tabular}{ccc}
\hline
$D$ & $d=0$ & $d=1$ \\ \hline
\multicolumn{1}{r}{4} & \multicolumn{1}{l}{24.086798714692429504} &
\multicolumn{1}{l}{24.361407377724659906} \\
\multicolumn{1}{r}{5} & \multicolumn{1}{l}{24.424509979572052554} &
\multicolumn{1}{l}{24.428687415859914431} \\
\multicolumn{1}{r}{6} & \multicolumn{1}{l}{24.429124533510868656} &
\multicolumn{1}{l}{24.429142207076630442} \\
\multicolumn{1}{r}{7} & \multicolumn{1}{l}{24.429143334496913081} &
\multicolumn{1}{l}{24.429143367079353985} \\
\multicolumn{1}{r}{8} & \multicolumn{1}{l}{24.429143368495260376} &
\multicolumn{1}{l}{24.429143368526373359} \\
\multicolumn{1}{r}{9} & \multicolumn{1}{l}{24.429143368527356872} &
\multicolumn{1}{l}{24.429143368527373946} \\
\multicolumn{1}{r}{10} & \multicolumn{1}{l}{24.429143368527374357} &
\multicolumn{1}{l}{24.429143368527374363} \\ \hline
\end{tabular}
\end{center}
\end{table}

\begin{table}[H]
\caption{Energy of the first excited state of the bounded oscillator with $%
R=1$}
\label{tab:table4}
\begin{center}
\begin{tabular}{ccc}
\hline
$D$ & $d=0$ & $d=1$ \\ \hline
\multicolumn{1}{r}{3} & \multicolumn{1}{l}{11.134110459186094281} &
\multicolumn{1}{l}{11.158800510018056742} \\
\multicolumn{1}{r}{4} & \multicolumn{1}{l}{11.161646658411540522} &
\multicolumn{1}{l}{11.161732939936591584} \\
\multicolumn{1}{r}{5} & \multicolumn{1}{l}{11.161737758990207575} &
\multicolumn{1}{l}{11.161737854670925624} \\
\multicolumn{1}{r}{6} & \multicolumn{1}{l}{11.161737857894423316} &
\multicolumn{1}{l}{11.161737857940626897} \\
\multicolumn{1}{r}{7} & \multicolumn{1}{l}{11.161737857941670476} &
\multicolumn{1}{l}{11.161737857941681848} \\
\multicolumn{1}{r}{8} & \multicolumn{1}{l}{11.161737857941682032} &
\multicolumn{1}{l}{11.161737857941682034} \\
\multicolumn{1}{r}{9} & \multicolumn{1}{l}{11.161737857941682034} &
\multicolumn{1}{l}{11.161737857941682034} \\
\multicolumn{1}{r}{10} & \multicolumn{1}{l}{11.161737857941682034} &
\multicolumn{1}{l}{11.161737857941682034} \\ \hline
\end{tabular}
\end{center}
\end{table}

\end{document}